\documentclass[letterpaper,10pt,twocolumn]{ASPE_ExtendedAbstract}
\usepackage{setspace}
\usepackage{color}
\usepackage{cite}
\usepackage{units}
\usepackage[pdftex]{graphicx}
\usepackage{float}
\usepackage[cmex10]{amsmath}
\usepackage{balance}
\usepackage{subfig}
\usepackage{times}
\usepackage{graphicx}
\usepackage{amsmath}
\usepackage{amssymb}
\intextsep=\lineskip

\title{FleXstage: Lightweight Magnetically Levitated Precision Stage with Over-Actuation towards High-Throughput IC Manufacturing\footnote{Accepted for ASPE 38th annual meeting}}
\author{Jingjie Wu and  Lei Zhou}
 
\affiliations{Department of Mechanical Engineering\\
University of Wisconsin-Madison, WI, USA\\
\\
}

\begin{document}
\maketitle 
\bibliographystyle{vancouverASPE}

\section*{Introduction}

\begin{figure}[t!]
\centering
\subfloat{
\includegraphics[trim={0mm 0mm 0mm 0mm},clip,width =0.8\columnwidth, keepaspectratio=true]{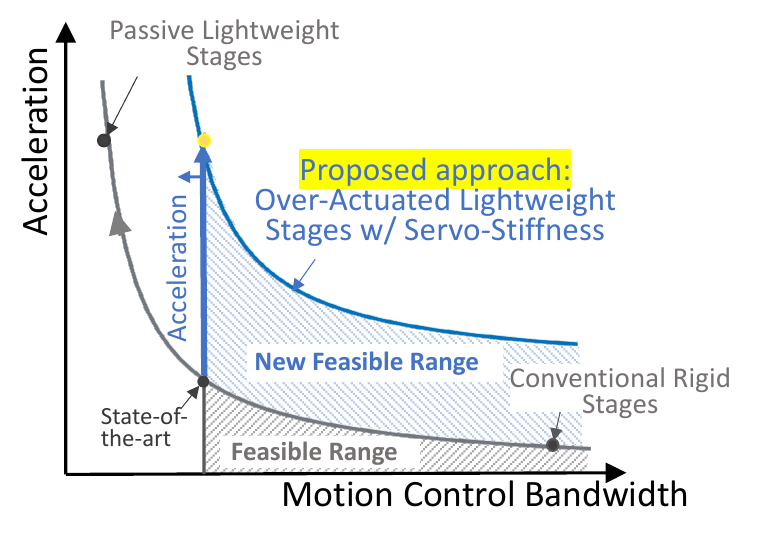}}
\vspace{-2mm}
\caption{Acceleration-bandwidth trade-off in today's precision positioning systems and motivation for the proposed method.}
\vspace{-2mm}
\label{fig:motivation}
% \end{figure}

% \begin{figure}[t!]
\centering
\subfloat{
\includegraphics[trim={0mm 0mm 0mm 0mm},clip,width =0.7\columnwidth, keepaspectratio=true]{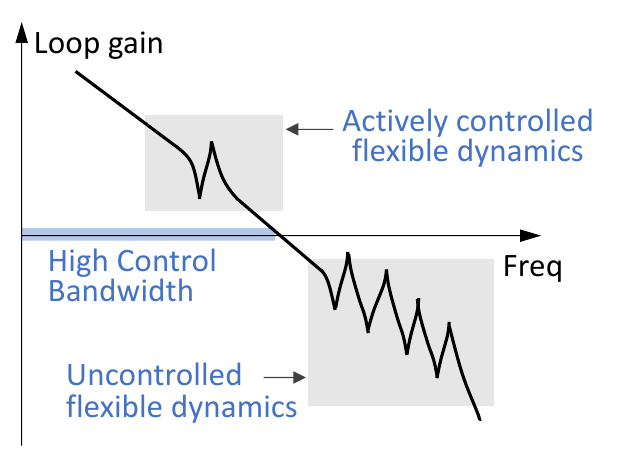}}
\vspace{-2mm}
\caption{Illustration of the  proposed lightweight stage design with active  flexible mode control.}
\vspace{-2mm}
\label{fig:proposed_cartoon}
\end{figure}

Precision motion stages play a critical role in various manufacturing and inspection equipment, for example, the wafer/reticle scanning in photolithography scanners and positioning stages in wafer inspection systems.
To meet the growing demand for higher throughput in chip manufacturing and inspection, it is critical to create new precision motion stages with higher acceleration capability with high control bandwidth \cite{oomen2013connecting}, which calls for the development of lightweight precision stages.  However,  in today's precision motion systems, only the rigid body motion of the system are under control, and the flexible dynamic systems are in open loop. For these systems, the motion control bandwidth is limited by the first structural resonance frequency of the stage, which enforces a fundamental trade-off between the stage's bandwidth and acceleration capability, as is shown in Fig.~\ref{fig:motivation}.

% current work
%In the past decades, a lot of research effort have studied  the design and control of lightweight precision motion stage to improve its overall performance. For example, Laro~et al.~\cite{laro2010design} presented an over-actuation approach to place actuators/sensors at the stage's nodal locations to prevent the flexible dynamics from being excited by the feedback loops. Van der Veen et al.~\cite{van2017integrating} studied the integrated topology and controller optimization for a simple 2D motion stage structure. Wu et al.~\cite{JingjieCoDesign} presented a nested CCD formulation of for lightweight precision stages with controller design constraints explicitly considered.
%Despite the advances, in these prior designs, the first resonance frequency of the  stage  sets an upper limit for the achievable control bandwidth, which enforces a fundamental trade-off between the stage's bandwidth and acceleration capability as shown in Fig.~\ref{fig:motivation}.

% describe algorithm and advantage
Aiming to overcome this trade-off, we have introduced a sequential structure and control design framework for lightweight stages with the low-frequency flexible modes of the stage are under active control \cite{JingjieASPE_2022}. To facilitate the controller design, we further propose to \textbf{minimize }the resonance frequency of the stage's mode being controlled and to  \textbf{maximize }the resonance frequency of the uncontrolled mode. The system's control bandwidth is placed in between the resonance frequencies, as shown in Fig.~\ref{fig:proposed_cartoon}. %The stage's geometry design optimization can be formulated as 
%\lei{add optimization problem here}
% \begin{align}  \label{eqn: shape_opt}
% \begin{split}
%     \min_{\theta_p}~~~&{J_m}(\theta_p),
% \\
%     \mathrm{s.t.}~~~&\omega_i  \leq \omega_{low}, ~~~~~~~ i=1,...,n
% \\
%     & \omega_j \geq \omega_{high}, ~~~~~ j=n+1,...,m 
% \\
%     & \theta_{p, min} \leq \theta_p \leq \theta_{p, max}, 
% \end{split}    
% \end{align}
% where the objective function $J_m$ represents the stage's weight, $\theta_p$ is a vector for the stage's geometric parameters,  $\omega_i$ is the $i$-th modal frequency with its corresponding vibration mode  actively controlled, and $\omega_j$ is the $j$-th resonance frequency where the corresponding mode shape is not controlled. $\omega_{low}$ is the upper bound for the actively-controlled resonance frequencies, and $\omega_{high}$ is the lower bound for the uncontrolled resonance frequencies. $\theta_{p, min}$ and $\theta_{p, max}$ are the lower and upper bounds for the stage's geometric parameter, respectively. 
Such an optimization process can enforce material removal in the stage's structure to allow for compliance in the actively-controlled flexible modes, and add material to stiffen the uncontrolled modes.

\begin{figure*}[t!]
\centering
\subfloat{
\includegraphics[trim={0mm 0mm 0mm 0mm},clip,width =1.9\columnwidth, keepaspectratio=true]{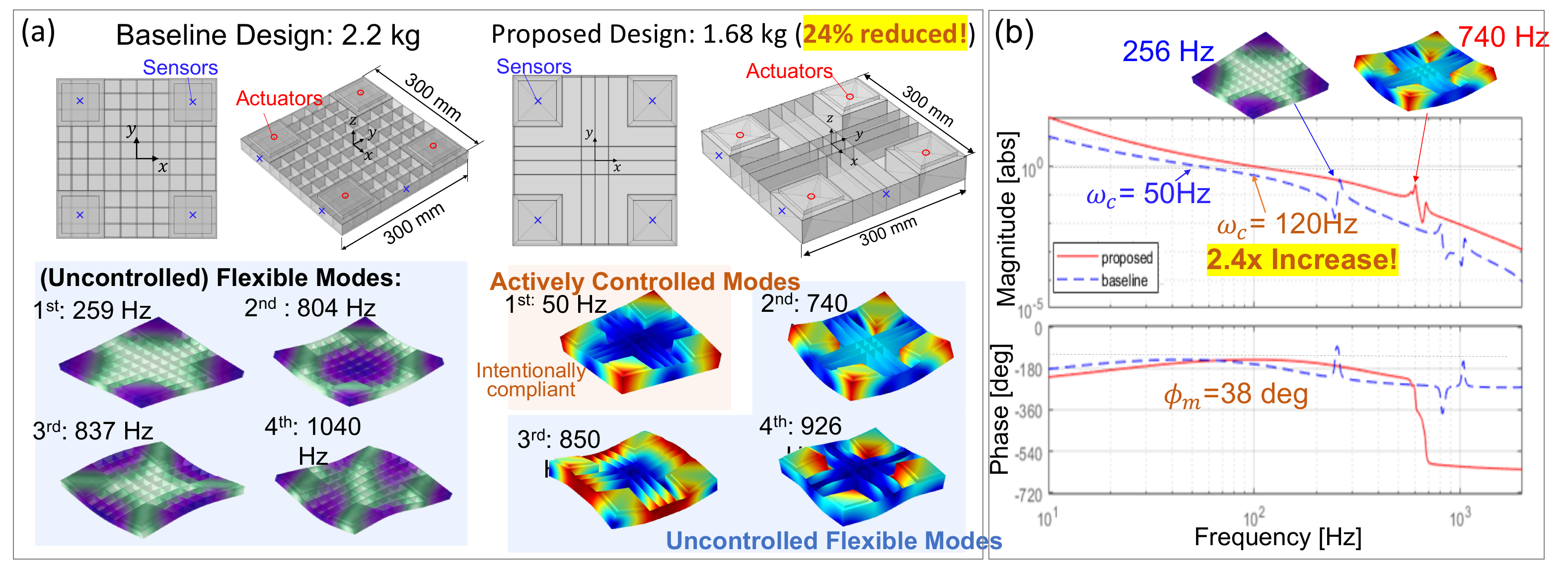}}
\vspace{-2mm}
\caption{(a) Proposed and baseline moving stages structure design. Note that the proposed case has one flexible mode with intentionally introduced compliance. (b) Comparison of loop gains of the proposed and baseline design in scanning direction. }
\vspace{-4mm}
\label{fig:case_2_definition}
\end{figure*}

% \begin{figure*}[t!]
% \centering
% \subfloat{
% \includegraphics[trim={0mm 0mm 0mm 0mm},clip,width =1.5\columnwidth, keepaspectratio=true]{Figures/Closed_loop.pdf}}
% \vspace{-2mm}
% \caption{Comparison of loop gains of the proposed design (red solid) and baseline design (blue dashed). (a) $y$-DOF (translation in the horizontal direction).  (b)  $z$-DOF (translation in the vertical direction). }
% \vspace{-3mm}
% \label{fig:case_2_closed_loop}
% \end{figure*}

% With the stage's structure designed, we further propose to use an optimization method to compute the optimal actuator/sensor placement. Our hypothesis is that maximizing the controllability/observability of the actively-controlled flexible modes while minimizing that of the uncontrolled modes will deliver the best positioning performance with reasonable control signal magnitude. Finally, the plant transfer function can be found after the actuator and sensor locations are optimized. Then, we decouple the system into 6 rigid-body dofs and 1 flexible mode under active control. For each DOF, a fixed-structure SISO PID controller with a second-order low-pass filter is designed to achieve the desired bandwidth. 

% describe simulation case
This abstract presents the design, optimization, building, and experimental evaluations for a lightweight magnetically levitated planar stage, which we call FleXstage, with first flexible mode actively controlled via over-actuation. The stage's structure is designed with intentionally introduced compliance to facilitate the flexible mode control and to reduce moving weight. 
Simulations show the proposed design is highly promising in enabling stages with lightweight without sacrificing control bandwidth. We have some preliminary results now and are still working on the experimental evaluations for the closed-loop system, and will present the results in the oral presentation.

\section*{Stage Design Optimization}
Figure.~\ref{fig:case_2_definition}a shows the structural design for the moving stage considered in this work, which is a rib-reinforced structure made of 7075-T6 aluminum alloy of 300~mm~$\times$~300~mm in size. There are four neodymium permanent magnet arrays of 70~mm $\times$ 70~mm $\times$ 6.35~mm arranged at the corners of the stage to provide both the thrust forces for planar motion and the levitation forces.  To reduce the stage's weight and have the first low-frequency flexible mode actively controlled, the stage's geometry can be determined by the following optimization problem: 
\vspace{-1mm}
%\lei{add optimization problem here}
\begin{align}  \label{eqn: shape_opt}
\begin{split}
    \min_{\theta_p}~~~&{J_m}(\theta_p),
\\
    \mathrm{s.t.}~~~&\omega_i  \leq \omega_{low}, ~~~~~~~ i=1,...,n
\\
    & \omega_j \geq \omega_{high}, ~~~~~ j=n+1,...,m 
\\
    & \theta_{p, min} \leq \theta_p \leq \theta_{p, max}.
\end{split}    
\end{align}
Here, the objective function $J_m$ is the stage's weight, $\theta_p$ represents the vector for the stage's geometric parameters,  $\omega_i$ is the $i$-th resonance frequency with its corresponding mode shapes actively controlled, and $\omega_j$ is the $j$-th modal frequency with corresponding vibration modes not controlled. $\omega_{low}$ is the upper bound for the actively-controlled modal frequencies, and $\omega_{high}$ is the lower bound for the uncontrolled modal frequencies. $\theta_{p, min}$ and $\theta_{p, max}$ are the lower and upper bounds for the stage's geometric parameter, respectively. By solving this optimization program, the material of the stage is partially removed to allow for compliance in the actively controlled flexible modes while the material kept can stiffen the uncontrolled modes.
Here, $\omega_{low}$ is chosen to be 50 Hz so that the control bandwidth can crossover well beyond it and $\omega_{high}$ is 560 Hz from sweeping to minimize the stage's weight and the limitation on achievable control bandwidth. 
The first four vibration mode shapes and corresponding resonance frequencies of the optimal stage structure are shown in Fig.~\ref{fig:case_2_definition}a. For comparison, a baseline stage structure in Fig.~\ref{fig:case_2_definition} is designed to constrain the first resonance frequency above 250~Hz with a target bandwidth of 50~Hz.

\section*{Actuator and Sensor Placement}

With the optimal stage's structure from \eqref{eqn: shape_opt}, the actuator and sensor placement optimization problem with first several flexible modes actively controlled  can be formulated as 
\vspace{-2mm}
\begin{align}  \label{eqn:act_opt}
\max_{\theta_a\in{D_a}}J_a(\theta_a)  = \sum_{i=1,...,n}W_{pi}(\theta_a) - \gamma\sum_{i=n+1,...,m} W_{pi}(\theta_a),
\end{align}
\vspace{-4mm}
\begin{align}  \label{eqn:sen_opt}
\max_{\theta_s\in{D_s}}J_o(\theta_s)  = \sum_{i=1,...,n}W_{oi}(\theta_s) - \gamma\sum_{i=n+1,...,m}W_{oi}(\theta_s),
\end{align}
\vspace{-6mm}

% \lei{put the optimization problem formulation here, and also include the discussion of changing gamma.}
where $\theta_a$ and $\theta_s$ are vectors of actuator and sensor placement parameters, respectively; $D_a$ and $D_s$ are the design feasible domains for actuator/sensor locations,  and $\gamma$ is a positive user-defined weighting constant. $W_{pi}$ and $W_{oi}$ are the controllability and observability grammians of $i$-th flexible mode, respectively, which can be calculated as 
\begin{align}
    W_{pi} = \frac{\| \phi_i(\theta_a)^\top B_a(\theta_a)\|_2^2}{4\zeta_i\omega_i}, \label{eq:W_p}\\
    W_{oi} = \frac{\| C_s(\theta_s)^\top \phi_i(\theta_s)\|_2^2}{4\zeta_i\omega_i}\label{eq:W_o},
\end{align}
\vspace{-6mm}

where $\phi_i$ is the mass-normalized mode shape of $i$-th flexible mode, $B_a$ and $C_s$ are the force and measurement assembling matrices, $\zeta_i$ is the modal damping ratio, and $\omega_i$ is the $i$-th resonance's natural frequency. The controllability/observability grammians $W_{pi}$ and $W_{oi}$ quantitatively represent the controllability/observability of the corresponding flexible mode, reflecting on its peak resonance magnitude in the system's frequency response.

With actuator/sensor placement optimization in \eqref{eqn:act_opt} and \eqref{eqn:sen_opt}, our goal is to maximize the controllability/observability for the actively-controlled modes to lower the required controller gain, and to minimize those of the uncontrolled modes to mitigate their coupling effect in the control systems and thus facilitate the controller design. It is worth noting such trade-off between the two design goals is determined by the value of weighting parameter $\gamma$ as: a low value in $\gamma$ emphasizes reducing the needed control effort for actively-controlled modes, and a high value in $\gamma$ emphasizes reducing the cross-talk between uncontrolled modes and controlled modes for higher control bandwidth. For the baseline case, we do not apply such placement optimization.

\begin{figure}[t!]
\centering
\subfloat{
\includegraphics[trim={0mm 0mm 0mm 0mm},clip,width =1\columnwidth, keepaspectratio=true]{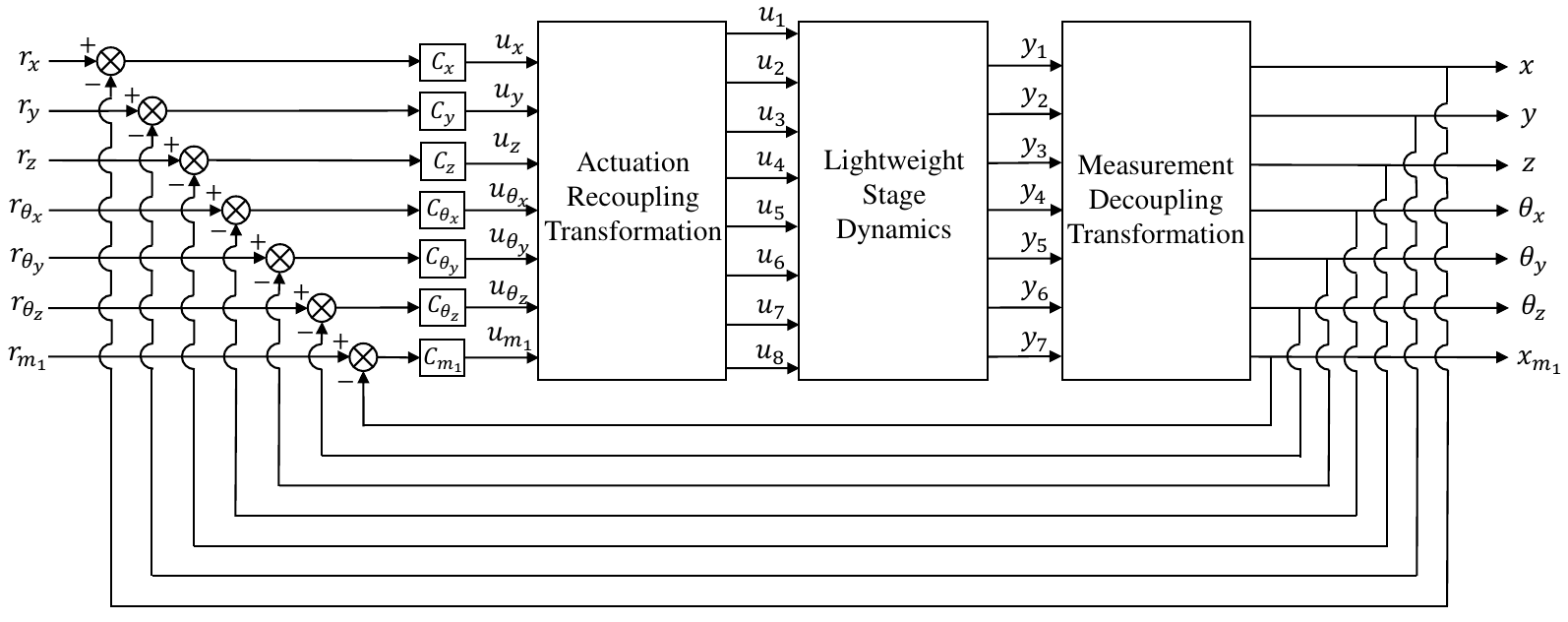}}
\vspace{-2mm}
\caption{Control block diagram for the lightweight precision positioning stage with model decoupling. }
\vspace{-5mm}
\label{fig:control_diagram}
\end{figure}

\section*{Feedback Control Design}

After the stage's hardware design (including both stage's structure and actuator/sensor placement) is optimized, the plant dynamic model can be derived. Finally,  feedback controllers are designed for each motion axis to attain the target closed-loop control performance.
Figure~\ref{fig:control_diagram} shows a block diagram for the closed-loop system with all six rigid-body DOFs and one flexible mode of the lightweight stage under active control. 
Here, the lightweight stage plant dynamics $P:u\to y$
can be obtained after solving  \eqref{eqn: shape_opt}, \eqref{eqn:act_opt}, and \eqref{eqn:sen_opt}. The actuator control inputs $u$ and the sensor measurements $y$ are recoupled and decoupled by transformation matrices, respectively, to obtain decoupled system DOFs. Seven single-input, single-output (SISO) feedback controllers can then be synthesized for the seven decoupled channels %: $u_z\to x_z$ , $u_{\theta x}\to \theta_x$, $u_{\theta y}\to \theta_y$ and $u_{m_1}\to q_1$, 
assuming the cross-talk between different channels is negligible near the target control bandwidth. For each DOF, a fixed-structure SISO controller is selected as \cite{franklin2002feedback}
\begin{align}
\begin{split}   \label{eqn:PID}
    C_k(s) = K_p\Big(\frac{s+\omega_i}{s}\Big)\Big(\frac{s}{\omega_d}+1\Big)\Big(\frac{\omega_{lp}^2}{s^2+2z_{lp}\omega_{lp}s+\omega_{lp}^2}\Big),
\end{split}
\end{align}
where the controller parameters are described in Table~\ref{table:PID_para}. This controller design significantly facilitates the parameter tuning process since all the controller parameters except the controller gain can all be determined by a single parameter, desired bandwidth $\omega_{bw}$ \cite{butler2011position}.
The proportional gain $K_p$ and the target bandwidth are chosen such that the control bandwidth is maximized for each channel while satisfying a robustness criteria\cite{ortega2004systematic}
\begin{align}  \label{eqn:robustness}
 \| S_k(s)\|_{\infty} \leq 2, k = 1, ..., n, 
\end{align}
where $S_k(s)$ is the closed-loop sensitivity function of the $k$-th channel as $S_k = (I-G_k C_k)^{-1}$. 
% % The controller design can be done by tuning the parameter $\omega_{bw}$ for each channel. 
% With the control effort signals $u_k$ for each channel computed, an actuation recoupling transformation is used to map the control signals to individual actuators. 

This PID controller structure is also used in the baseline case and tuned to achieve its target bandwidth. Fig.~\ref{fig:case_2_definition}b illustrates the loop gains of both proposed and baseline designs in $y$-DOFs, the performance of which are most critical. It can be observed that with sufficient stability margins in both cases, the control bandwidth in proposed design is significantly larger than the baseline design. The weight of the proposed stage design (1.68~kg) is reduced by 24\% compared to baseline design (2.21~kg). The comparisons demonstrate the potential of the proposed framework to improve the stage's acceleration capability while maintaining high bandwidth and robustness.

\begin{table}[t] 
\centering
\vspace{2mm}
\caption{Controller parameters  \cite{butler2011position}. }
\vspace{-6mm}
\label{table:PID_para}
    \begin{center} \begin{small}
        \begin{tabular}{ p{1.2cm} p{4cm}  p{1cm} }
        \hline
        Parameter & Description & Typical Value \\
        \hline
        $\omega_{bw}$  & Desired bandwidth [rad/s]  & --  \\
        
        % $m$    &  Stage mass [\rm{kg}]   & --  \\
        
        % $J$ & Stage moment of inertia [$\rm{kg\cdot m^2}$]  &  -- \\
        
        $\alpha$ & PID frequency ratio  &   3 \\
        
        $K_p$  &    Proportional gain  & --   \\
        
        $\omega_i$  & Integrator frequency  & $  \omega_{bw}/ \alpha^2$  \\
        
        $\omega_d$  & Differentiator frequency  & $\omega_{bw}/\alpha$ \\
        
        $\omega_{lp}$  & Lowpass filter frequency &  $  \alpha \omega_{bw}$  \\
        
        $z_{lp}$  & Lowpass filter damping ratio  & 0.7\\
        \hline

        \end{tabular}
    \end{small} \end{center}
\vspace{-6mm}
\end{table}

\section*{Prototype Introduction}

\begin{figure}[t!]
\centering
\subfloat{
\includegraphics[trim={0mm 0mm 0mm 0mm},clip,width =1\columnwidth, keepaspectratio=true]{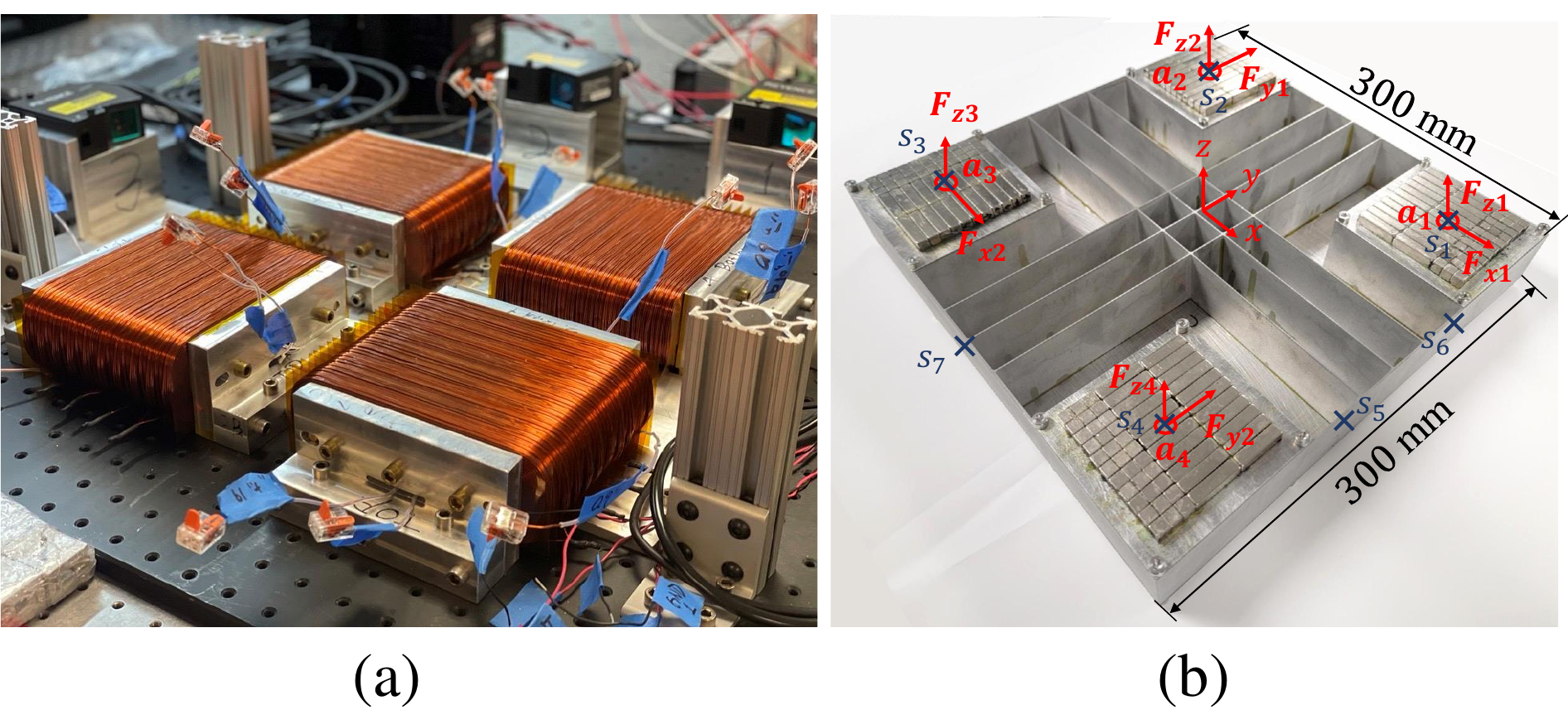}}
\vspace{-2mm}
\caption{(a) Stator assembly (b) Fabricated stage with forces location and directions labeled }
\vspace{-2mm}
\label{fig:stage_hardware}
\end{figure}

% \begin{figure}[t!]
% \centering
% \subfloat{
% \includegraphics[trim={0mm 0mm 0mm 0mm},clip,width =0.8\columnwidth, keepaspectratio=true]{Figures/Stator.jpg}}
% \vspace{-2mm}
% \caption{Planar motor stator.}
% \vspace{-2mm}
% \label{fig:stator}
% \end{figure}

\begin{figure}[t!]
\centering
\subfloat{
\includegraphics[trim={0mm 0mm 0mm 0mm},clip,width =1\columnwidth, keepaspectratio=true]{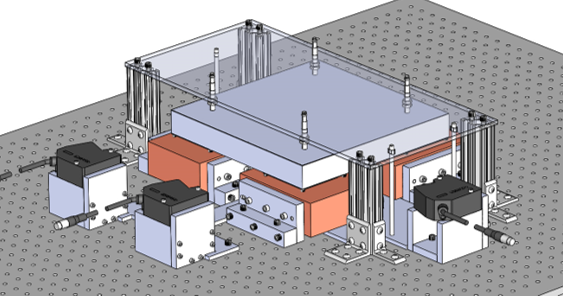}}
\vspace{-2mm}
\caption{Planar motor hardware CAD model for experimental evaluation. }
\vspace{-2mm}
\label{fig:motor_assembly}
\end{figure}

% \begin{figure}[t!]
% \centering
% \subfloat{
% \includegraphics[trim={0mm 0mm 0mm 0mm},clip,width =1\columnwidth, keepaspectratio=true]{Figures/Whole_setup.jpg}}
% \vspace{-2mm}
% \caption{Planar motor hardware setup. }
% \vspace{-2mm}
% \label{fig:setup}
% \end{figure}

\begin{figure}[t!]
\centering
\subfloat{
\includegraphics[trim={0mm 0mm 0mm 0mm},clip,width =1\columnwidth, keepaspectratio=true]{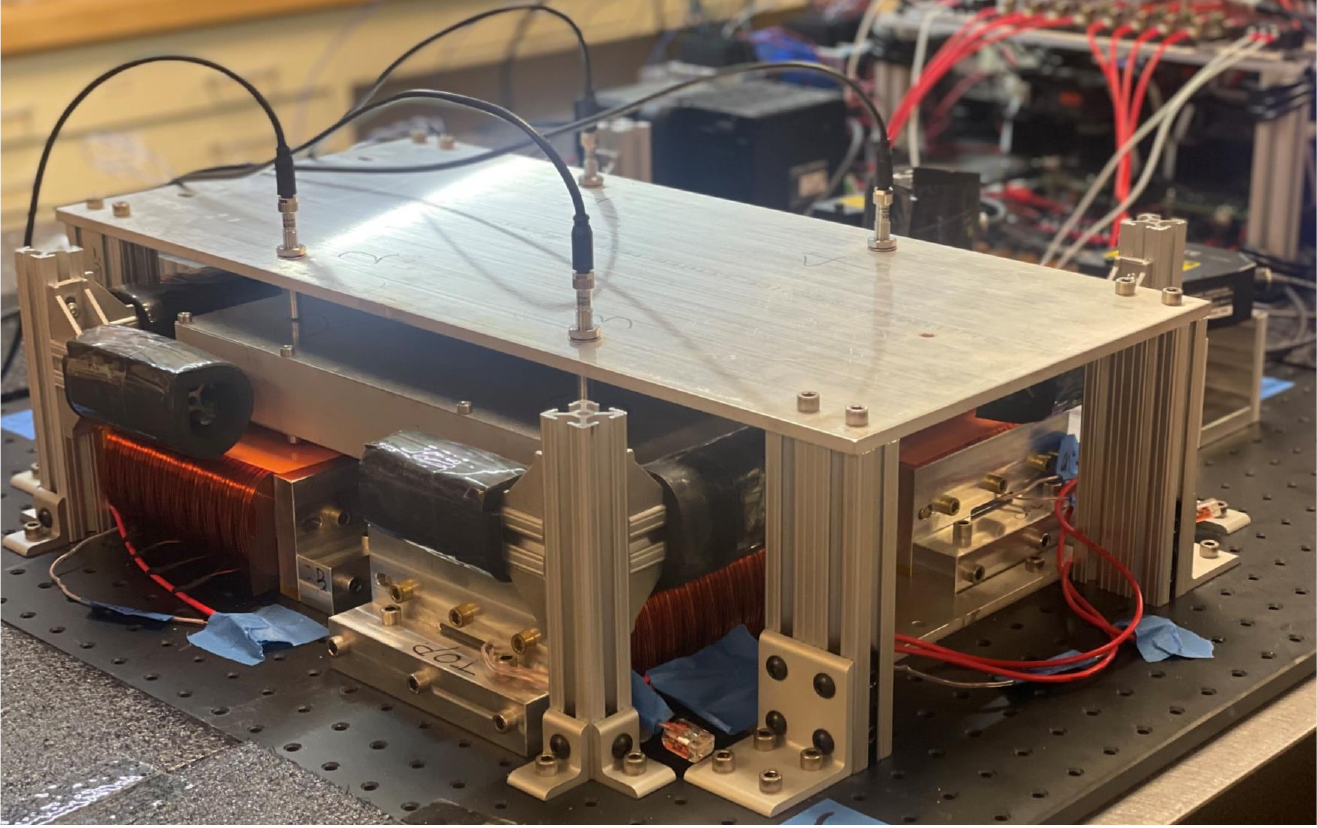}}
\vspace{-2mm}
\caption{Experimental Setup.}
\vspace{-2mm}
\label{fig:whole_setup}
\end{figure}

To verify the improvement in overall system performance, the optimized stage is fabricated as shown in Fig.~\ref{fig:stage_hardware}b. The magnet arrays at corners to provide thrust and levitation force are made of N52 NdFeB and have the pattern of a traditional 4-segment-per-spatial-period Halbach array, which can generate a higher magnetic field. Fig.~\ref{fig:stage_hardware}a shows the assembly of the four stators holding coil windings with the thickness being 15 mm according to parameter sweeping in the Ansys Maxwell electromagnetic simulations to maximize the instantaneous force capability and therefore the acceleration capability. The APEX PA12 power op-amp is used inside of a customized current controller with a bandwidth of around 5kHz to provide the current for the coils up to 10~A. Each coil consists of 100 turns of AWG\#19 to have sufficiently small resistance and thus provide the desired peak current. The overall planar motor CAD design for actuation and sensing of the stage is shown in Fig.~\ref{fig:motor_assembly} including four DW-AS-509-M8-390 inductive sensors to measure the vertical displacement and three LK-H152 laser displacement sensors to measure full x and y axis motion. The experimental setup of the whole planar motor is shown in Fig.~\ref{fig:whole_setup}.

\begin{figure}[t!]
\centering
\subfloat{
\includegraphics[trim={0mm 0mm 0mm 0mm},clip,width =1\columnwidth, keepaspectratio=true]{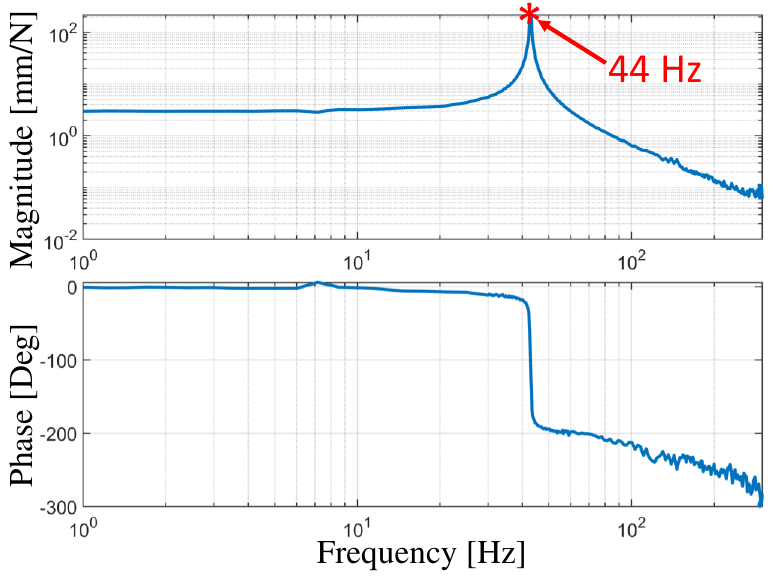}}
\vspace{-2mm}
\caption{Identified 7th channel plant frequency response.}
\vspace{-2mm}
\label{fig:Q}
\end{figure}

As a primary test of the planar motor system, we employ a lead-lag form PID controller for each motion axis and then tune all the controller parameters to only stabilize the closed-loop system. Then we perform a system identification for each DOF's transfer function by Dynamic Signal Analyzer to validate the stage's mass, mode shapes, and resonance frequencies derived from COMSOL Finite Element simulation. Fig.~\ref{fig:Q} shows the identified frequency response of the 7th channel for the first mode being actively controlled. The 1st resonance frequency of the fabricated stage structure is 44Hz, which well matches with our prediction. We are currently working on further system identification and tune the controllers to obtain a high control bandwidth, and the results will be presented at the conference and our future works.

%%%%%%%%%%%%%%%%%%%%%%%%%%%%%%%%%%%%%%%%%%%%%%%%%%%

\vspace{-2mm}

\balance
\bibliography{Co_design_ASPE}

\end{document}